\documentclass[10pt,conference]{IEEEtran}
\IEEEoverridecommandlockouts

\usepackage{cite} 

\usepackage{amsmath,amssymb,amsfonts}
\usepackage{algorithmic}
\usepackage{graphicx}
\usepackage{textcomp}
\def\BibTeX{{\rm B\kern-.05em{\sc i\kern-.025em b}\kern-.08em
    T\kern-.1667em\lower.7ex\hbox{E}\kern-.125emX}}
\def\SymboleoNLP/{\textsc{SymboleoNLP}}
\def\SymboleoPC/{\textsc{SymboleoPC}}
\def\SymboleoWeb/{\textsc{SymboleoWeb}}
\def\SymboleoSC/{\textsc{Symboleo2SC}}
\def\SymboleoJS/{\textsc{SymboleoJS}}
\def\SymboleoAC/{\textsc{SymboleoAC}}

\usepackage{wrapfig}
\usepackage{orcidlink}
\usepackage{doi}
\usepackage{url}
\usepackage{xspace}
\usepackage{flushend} 
\usepackage{lipsum} 
\usepackage{colortbl}
\usepackage{comment}
\usepackage{adjustbox}
\usepackage[T1]{fontenc}
\usepackage{array}
\usepackage{rotating}
\usepackage{multirow}
\usepackage{graphicx}
\usepackage{makecell}
\usepackage[utf8]{inputenc}
\usepackage{longtable,tabularx,ltxtable,ragged2e}
\usepackage{booktabs}   
\usepackage{ltablex}
\usepackage{xspace}
\usepackage{acronym}
\usepackage{rotating}
\usepackage[inline]{enumitem}
\usepackage{makecell}
\usepackage{hyperref}
\hypersetup{colorlinks,allcolors=black} 
\usepackage{orcidlink}
\usepackage[scaled=0.86]{helvet}
\usepackage{footmisc}

\usepackage{colortbl}
\PassOptionsToPackage{table,xcdraw}{xcolor}

 \definecolor{lightgreen}{rgb}{0.88, 1.0, 0.88} 

\usepackage[framemethod=TikZ]{mdframed}
\usepackage[frozencache=true,cachedir=minted-cache]{minted} 

\definecolor{light-yellow}{rgb}{0.99, 0.99, 0.89}

\hypersetup{
    colorlinks=true,
    linkcolor=black,
    filecolor=magenta,      
    urlcolor=blue,
    citecolor=black,
    pdfborder={0 0 0}
}

\mdfdefinestyle{MyFrame}{%
    linecolor=black,
    outerlinewidth=0.15pt,
    roundcorner=3pt,
    innertopmargin=2pt,
    innerbottommargin=2pt,
    innerrightmargin=4pt,
    innerleftmargin=4pt,
    backgroundcolor=light-yellow}
\mdfdefinestyle{MyFrameSimple}{%
    linecolor=black,
    outerlinewidth=0.15pt,
    roundcorner=3pt,
    innertopmargin=2pt,
    innerbottommargin=2pt,
    innerrightmargin=4pt,
    innerleftmargin=4pt}
\usepackage[scaled=0.86]{helvet}

\usepackage{listings}
\AtBeginDocument{\counterwithout{lstlisting}{section}}

\lstdefinelanguage{symboleo}{
language=C++,
morekeywords={Domain, endDomain, Contract, Declarations, Preconditions,
Postconditions, Obligations, Surviving, Powers, Power, Obligation, Number,
Constraints, endContract, isA, isAn, Enumeration, with, String, Date, Boolean,
Asset, Event, Role, Env, or, and, not, true, false, O, P, Suspended, Resumed, Discharged, Terminated,
Happens, HappensBefore, HappensAfter, HappensWithin, Occurs, IsEqual, IsOwner, CannotBeAssigned,
Triggered, Activated, Exerted, Expired, Fulfilled, Discharged, Violated, FulfilledObligations,
RevokedParty, AssignedParty, Rescinded, Create, UnsuccessfulTermination, Active, InEffect, 
Suspension, SuccessfulTermination, Discharge, Violation, Fulfillment, UnsuccessfulTermination, 
String.concat, Math.pow, Math.abs, WhappensBefore, ShappensBefore, Assign,, HappensAssign, self,
UnAssign, Form, Rescission, read, write, Grant, Revoke, To, On, by, DataTransfer, Resource, thirdParty},
morecomment=[l]{--},
    backgroundcolor=\color{CLightGray},
    keywordstyle=\color{blue!80}\bfseries,
    basicstyle=\tiny\ttfamily,
    commentstyle=\color{green!40!black},,
    numbers=left,
    numberstyle=\tiny\ttfamily,
    numbersep=1em,
    showstringspaces=false,
    breaklines=true,
    frame=lines,
}

\definecolor{codegreen}{rgb}{0,0.4,0}
\definecolor{codegray}{rgb}{0.5,0.5,0.5}
\definecolor{codepurple}{rgb}{0.58,0,0.82}
\definecolor{backcolour}{rgb}{0.95,0.95,0.92}
\definecolor{CLightGray}{gray}{0.9}
\definecolor{lightyellow}{rgb}{0.98, 0.98, 0.82}

\lstdefinestyle{mystyle}{
    backgroundcolor=\color{backcolour},   
    commentstyle=\color{codegreen},
    keywordstyle=\color{magenta},
    numberstyle=\tiny\color{codegray},
    stringstyle=\color{codepurple},
    basicstyle=\ttfamily\tiny,
    breakatwhitespace=false,         
    breaklines=true,                 
    captionpos=b,                    
    keepspaces=true,                 
    numbers=left,                    
    showspaces=false,                
    showstringspaces=false,
    showtabs=false,                  
    tabsize=2,
    frame=lines,
    framexleftmargin=0.9em,
    xleftmargin=1em,
    numbersep=0.5em
}

\begin{document}

\newcommand{\email}[1]{\texttt{#1}}
\def \Symboleo/{\textsc{Symboleo}}



\title{Towards the LLM-Based Generation of Formal Specifications from Natural-Language Contracts: Early Experiments with \Symboleo/\\
\thanks{This work is funded by the ORF-RE Grant \textit{CyPreSS: Software Techniques for the Engineering of Cyber-Physical Systems}, the SSHRC Grant \textit{Autonomy through Cyberjustice Technologies (ACT)}, and the NSERC Discovery Grant \textit{From Legal Contracts to Smart Contract}.}
}

\author{\IEEEauthorblockN{Mounira Nihad Zitouni, Amal Ahmed Anda\orcidlink{0000-0001-7851-0199}, Sahil Rajpal, Daniel Amyot\orcidlink{0000-0003-2414-1791}, John Mylopoulos\orcidlink{0000-0001-8926-5221}}\\

\IEEEauthorblockA{\textit{School of EECS, University of Ottawa, Ottawa, K1N 6N5, Canada}\\ 
\email{\{mzito042, aanda, srajp029, damyot, jmylopou\}@uottawa.ca}\\
\url{https://sites.google.com/uottawa.ca/csmlab/} 
}

}
\maketitle

\begin{abstract}
Over the past decade, different domain-specific languages (DSLs) were proposed to formally specify requirements stated in legal contracts, mainly for analysis but also for code generation. \Symboleo/ is a promising language in that area. However, writing formal specifications from natural-language contracts is a complex task, especial for legal experts who do not have formal language expertise. This paper reports on an exploratory experiment targeting the automated generation of \Symboleo/ specifications from business contracts in English using Large Language Models (LLMs). Combinations (38) of prompt components are investigated (with/without the grammar, semantics explanations, 0 to 3 examples, and emotional prompts), mainly on GPT-4o but also to a lesser extent on 4 other LLMs. The generated specifications are manually assessed against 16 error types grouped into 3 severity levels. Early results on all LLMs show promising outcomes (even for a little-known DSL) that will likely accelerate the specification of legal contracts. However, several observed issues, especially around grammar/syntax adherence and  environment variable identification (49\%), suggest many areas where potential improvements should be investigated. 
\end{abstract}

\begin{IEEEkeywords}
Code Generation, DSL, Large Language Model, Legal Contract, Requirements Specification, \textsc{Symboleo}.
\end{IEEEkeywords}

\section{Introduction}
\label{sec:Introduction}

Legal contracts are agreements between parties that delineate the obligations, rights, and conditions that must be adhered to. They also specify requirements on parties that must be monitored for violations and other types of breaches. As such contracts are most often written in natural language, they are subject to several types of ambiguities and inconsistencies that are typically difficult to analyze automatically. 

Several Domain-Specific Languages (DSLs) enable the formal specification and analysis of contracts~\cite{Soavi2022Review}. For example, \Symboleo/ was recently proposed as a structured and machine-readable language for specifying contractual obligations, powers, and domain concepts, while enabling design-time verification of logical properties, as well as code generation of smart contracts that monitor contract executions~\cite{parvizimosaed2022specification}.

However, writing contract specifications requires skills that contract authors (e.g., lawyers) and parties (e.g., buyers and sellers) typically do not master. Even for DSL experts, this conversion takes time and is prone to errors as it involves handling  complex DSL constructs and legal interpretations. 

Several approaches, typically based on Natural Language Processing (NLP) or Machine Learning (ML), have been proposed to support the automatic generation of specifications and code in a given DSL from natural-language documents, including from contracts (e.g., \cite{Soavi2022Review,Ge2023-ContractsMTL,FANTONI2021-Contract-NLP}). For \Symboleo/ in particular, Soavi et al.~\cite{Soavi2022ContrattoA} have explored tool-supported semantic annotations without full specification generation, whereas Meloche et al.~\cite{meloche2023towards,SymboleoWebaper2024} have explored full generation, but limited to specific refinements of contract templates.

In recent years, Large Language Models (LLMs) such as GPT~\cite{ChatGPT} have also been investigated in the context of generating specifications in a DSL from natural-language documents~\cite{Lamas2024DSLgen,bassamzadeh2024DSLcodeGen,Mosthaf2024-DSLassistant,chen2023gpt4goalmodels}. In the legal field, where documents are typically long, intricate, and written using highly technical terminology, tasks such as legal contract specification and analysis may be more productive with the use of LLMs~\cite{hassani2024rethinking}. 

This paper aims to investigate the potential and limitations of LLMs in automatically generating formal specifications from natural-language legal business contracts. The focus is on one particular DSL, \Symboleo/, as this language supports downstream software engineering tasks such as verification through model checking~\cite{parvizimosaed2024symboleopc} and smart contract code generation for the Hyperledger Fabric blockchain platform~\cite{Rasti2024-SoSyM}.

A key strategy to good LLM performance for such a task is prompt engineering~\cite{Sahoo2024-SurveyPromptEngineering}, whereby LLMs can be guided in generating outputs that reflect the required format and structure, especially for target DSLs that are typically unknown to the LLMs involved. Our main research question (\textbf{RQ}) here is:

\begin{mdframed}
\textbf{RQ}: What combination of prompting strategies enables the most accurate generation of \Symboleo/ specifications from natural-language business legal contracts?
\end{mdframed}

As \Symboleo/ is a new DSL and not yet commonly used, this first exploratory experiment will focus on combinations of four prompt components, namely with/without i)~the DSL grammar, ii)~explanations of core DSL semantics, iii)~zero, one, and few shots, and iv)~emotional prompting. A total of 38 combinations are tested on ChatGPT-4o~\cite{ChatGPT}, but two extreme combinations are also tested on other LLMs that have simple Web-based access, namely Llama 3.2\cite{Llama}, Claude 3.5 Haiku~\cite{Claude}, Mistral 7B~\cite{Mistral}, and Gemini 1.5 Pro 002~\cite{Gemini}. 

The evaluation of the resulting specifications involves counting errors in the generated specifications along 16 types grouped into 3 severity levels. Although we expect all specifications to contain errors, we are interested in characterizing them so better approaches can mitigate in the future. We also appreciate that it is usually far simpler and faster to fix an incorrect specification than it is to create one from scratch, especially for non-experts.

The rest of this paper is structured as follows. Section~\ref{sec:Background} provides necessary background on \Symboleo/. Section~\ref{sec:PromptEngineering} presents the experiment in terms of prompt combinations, while Section~\ref{sec:Evaluation} gives the evaluation results along our error-related metrics. Sections~\ref{sec:Discussion}, \ref{sec:RelatedWork}, and \ref{sec:Conclusion} respectively provide a discussion, related work, and the conclusion.

\section{Background}
\label{sec:Background}

\Symboleo/ is an ontology-based formal specification language designed to enable the creation of legally accurate and executable specifications of contracts, reduce ambiguity, and make contract compliance verifiable and enforceable through automated smart contract systems~\cite{parvizimosaed2022specification}.  

\subsection{\Symboleo/ Ontology}
\Symboleo/ is rooted in a strong ontology based on legal theories, particularly the Unified Foundation Ontology (UFO) and its legal extension, UFO-L\cite{griffo2015towards}. UFO-L provides basic legal concepts (e.g., roles, assets, obligations, powers, and events)  essential for defining contractual relationships~\cite{sharifi2023towards,parvizimosaed2022specification}, which are adapted in \Symboleo/'s ontology.

\textbf{Roles:} Roles define the participants (e.g., buyer, seller) and their responsibilities within the contract.

\textbf{Assets:} Assets represent items of value involved in the contract (e.g., goods or services). They are defined with specific properties like quality and quantity, ensuring clarity in what is being exchanged.

\textbf{Obligations:} Obligations indicate duties one role owes another, triggered by specific conditions. Each obligation has an antecedent and a consequent legal position. An obligation can, for instance, mandate that a buyer pay by a particular date or that a seller provides products on a specific date. 

\textbf{Powers:} Powers are rights that allow a role to instantiate, change, suspend, or terminate obligations when conditions are met. Powers provide flexibility by enabling parties to adjust obligations or impose penalties in response to delays or breaches of the contract.

\textbf{Event}: Event (based on event calculus) occur at a time point (e.g., a date/time), and cannot change. Events have pre- and post-state situations. For example, \textit{delivered} is an event whose pre-state is `being in transit' and post-state is `being
at the point of destination'. Events have typed parameters, some of which with values that can come from the environment.

The full ontology additionally includes concepts for contracts, parties, situations, time points, and time intervals~\cite{parvizimosaed2022specification}.

\subsection{Grammar}

\Symboleo/'s grammar\footnote{\Symboleo/ grammar in Xtext: \url{ https://bit.ly/Symboleo-Xtext}\label{foot:xtext}} is defined using Xtext~\cite{bettini2016Xtext} and contains about 70 rules. The first part of a specification contains the contract's \textit{domain} model, which extends the ontology concepts (mainly assets, roles, and events) describing contract-specific concepts and their attributes.

A \textit{contract signature} then defines the name of the contract together with its typed parameters (e.g., roles, quantities, deadlines). In essence, a \Symboleo/ specification is a contract template whose parameters must be provided with specific values during instantiation. 

A specification then describes \textit{declarations}, which initialize local variables with initial values. \Symboleo/ supports basic types (e.g., Boolean, Date, Number, String, enumerations...) and user-defined types declared in the domain models.

\begin{lstlisting}[language=symboleo, basicstyle=\footnotesize\ttfamily, numbers=none,  columns=fullflexible]
Domain MyDomain 
  Seller isA Role name: String;
  Deliv isAn Event with Env qty: Number; // ...
endDomain 
Contract MyContract (s : Seller, name:String, ...)
  Declarations
    b: Buyer with buyername:=name;  // ...
  Obligations
    delivery: Obligation(seller, buyer, true, 
      WhappensBefore(deliv, dueDate)); //...
  Powers
    suspendDelivery : Happens(Violated(obligations.payment)) -> Power(s, b, true, Suspended(obligations.delivery)); //...
  Constraints
    not(IsEqual(s, b)); // seller <> buyer
\end{lstlisting}

Optionally, \textit{preconditions} confirm correct role assignment, asset definition, and initial parameters, while \textit{postconditions} define expected outcomes upon contract termination.

The key legal elements in a specification are \textit{obligations} and \textit{powers}. They are formalized using a syntax that specifies antecedents and consequents akin to an ``if-then'' format. Obligations dictate duties, while powers allow parties to adjust contract states based on specified conditions. For instance, an obligation might require ``the seller to deliver goods to the buyer by a set date'' whereas a power might allow the buyer to suspend delivery if payment is not on its way~\cite{parvizimosaed2022specification}.

Finally (and optionally), \textit{constraints} enforce certain rules within the contract (e.g., the seller and buyer are different). 

\section{Prompt Engineering }
\label{sec:PromptEngineering}
This section presents the selected components that can be assembled to form the prompts used in this experiment. The prompts, corresponding \Symboleo/ specifications, and analysis results are freely available online~\cite{zitouni_2024}. Our preliminary experiment exclusively focuses on prompts for out-of-the-box Web-based LLM environments. LLM parameter tuning (e.g., temperature) and fine-tuning (which would require many small code generation examples that we do not have) are particularly outside the scope of our study. 

\subsection{Prompt Components}
Sahoo et al.~\cite{Sahoo2024-SurveyPromptEngineering} have reviewed many prompt engineering techniques, including context provision (e.g., DSL grammar and semantics), few-shot learning, and emotional prompting. These are the building blocs explored here.

\subsubsection{Xtext Grammar Explanation (Syntactical Context)}
We used the syntax definition of \Symboleo/'s in Xtext\footref{foot:xtext} to guide the model to conform to the stringent syntactical grammar of the DSL. The grammar also provides the model with a foundational template through this organized syntactic format, facilitating its understanding of \Symboleo/'s elements, such as obligations, roles, powers, and constraints. This grammar serves as a syntactic framework, which is likely beneficial for generating compliant specifications in domains that require precise structural adherence. The prompt is as follows (the grammar, not shown here, is the one available online\footref{foot:xtext}):

\begin{mdframed}[style=MyFrame]\small
\textsf{Here is Symboleo's syntax in Xtext format:}
\end{mdframed}

\subsubsection{Modalities and Events (Theoretical Context)}
Legal contracts encompass intricate modalities in which obligations and powers are contingent on specific trigger conditions and outcomes. To the LLM about these aspects, we provided additional context in select prompts, detailing \Symboleo/'s semantics for interpreting obligations, powers, and various non-trivial functions for observing events. This may enable the model to more effectively interpret events and triggers within contract clauses~\cite{parvizimosaed2022specification,meloche2023towards}.

\begin{mdframed}[style=MyFrame]\small
\textsf{Also, please note that in Symboleo:\\
- Obligations have the format "Oid: [trigger ->] O(debtor, creditor, antecedent, consequent)" where creditor and debtor are roles whereas the trigger, antecedent, and consequent are legal situations defined by propositions.\\
- Surviving obligations are the obligations that remain in effect after the termination of a contract.\\
- Powers are specified as "Pid: [trigger ->] P(creditor, debtor, antecedent, consequent)" are used to create, change or terminate an obligation or another power.\\
\\
Regarding Symboleo's semantics, please note that:\\
- Happens(e1) is true if event e1 has happened.\\
- HappensAfter(e1, p1/e2) is true if event e1 happened after time point p1 or event e2.\\
- WhappensBefore(e1, p1/e2) is true if event e1 happened before time point p1 or event e2.\\
- ShappensBefore(e1, e2) is true if events e1 and e2 have happened and e1 happened before e2.\\
- HappensWithin(e1, int1) is true if event e1 happened within interval int1 (where an interval consists of two time points).\\
- HappensWithin(e1, sit1) is true if event e1 happened when situation sit1 was held (e.g., an obligation is in violation state).}\\
\end{mdframed}

\subsubsection{Emotional Prompting (Motivational Directive)}
This study also briefly investigates emotional directives to determine whether the affective language in prompts affects the engagement or output quality of the LLM. Although this method is somewhat unconventional, it is consistent with findings in prompt engineering research, which indicate that motivational prompts can occasionally improve the focus and engagement of outputs~\cite{hassani2024rethinking}.

\begin{mdframed}[style=MyFrame]\small
\textsf{Do a good job as this is the most crucial point in my dream career and everything is relying upon it.}
\end{mdframed}

\subsubsection{Example Scenarios (Few-Shot Learning)}
Examples may also be used in prompt engineering, as the LLM may benefit from exposure to similar tasks or examples before generating a response. We provided example scenarios in various combinations to evaluate their effect on the model's ability to generate accurate \Symboleo/ specifications. These scenarios included legal contract examples with their corresponding \Symboleo/ specifications. By testing prompts with zero to three examples, we can investigate the influence of example presence, diversity, and ordering on enhancing the LLM's performance in producing precise contract specifications. Each prompt may include 0 to 3 examples, for example:
\begin{mdframed}[style=MyFrame]\small
\textsf{Here is the First example of a legal contract in natural language, followed by its Symboleo specification:\\
... \textit{[Natural-language contract]}\\
The corresponding Symboleo specification is:\\
... \textit{[Symboleo code]}\\
Second example of a legal contract in natural language, followed by its Symboleo specification:\\
... \textit{[Natural-language contract]}\\
The corresponding Symboleo specification is:\\
... \textit{[Symboleo code]}\\
Third example of a legal contract in natural language, followed by its Symboleo specification:\\
... \textit{[Natural-language contract]}\\
The corresponding Symboleo specification is:\\
... \textit{[Symboleo code]}}\\
\end{mdframed}

There are 3 specific scenarios, based on existing contracts:

\textbf{Scenario A: Medical Supply Contract}~\cite{Rasti2024-SoSyM}:
This contract (10 clauses) outlines the terms between a medical consortium (MCDC) and Pfizer for the production and delivery of vaccines. Pfizer must deliver vaccine doses according to specified timelines, meet dosage requirements, and ensure appropriate storage conditions, while notifying MCDC of lead times. Payment is based on actual quantities delivered, with specific processing timelines. MCDC is restricted from issuing stop-work orders unless legally required, and Pfizer is exempt from liability for delays caused by regulatory or manufacturing hurdles. The corresponding \Symboleo/ specification has 97 lines, with 11 domain elements, 3 obligations, and 3 powers.

\textbf{Scenario B: Energy Supply Agreement}~\cite{parvizimosaed2022specification}:
This agreement (7 clauses) governs the relationship between a Distributed Energy Resource Provider (DERP) and the California Independent System Operator (CAISO) for energy delivery and payments. DERP must fulfill energy delivery obligations as per agreed bids, while CAISO is required to process payments within specified timeframes. CAISO can terminate the agreement for delivery failures or unpaid invoices, and DERP may terminate with 90 days' notice. Penalties can be imposed on DERP for failure to meet energy commitments. Its \Symboleo/ specification contains 65 lines, including 18 domain elements, 8 obligations, and 2 powers.

\textbf{Scenario C: Meat Sale Contract}~\cite{parvizimosaed2022specification}:
This contract (8 clauses) specifies an international transaction between a seller and a buyer for the delivery of a specified quantity and quality of meat. The seller must deliver the meat within a set period, and the buyer must complete payment by the agreed date, with penalties for late payment. The buyer is granted termination rights if delivery is delayed beyond a 10-day grace period. Both parties are bound by a confidentiality clause that applies during and after the contract's execution. The corresponding \Symboleo/ specification contains 56 lines, including 10 domain elements, 3 obligations, and 3 powers.

\subsection{Test Cases}
To assess the impact of prompt engineering on \Symboleo/ specification generation, we examine various combinations of prompt components, each designed to guide the model in a unique way. The following is a summary of the 38 prompt test cases, which all start with this initial statement:
\begin{mdframed}[style=MyFrame]\small
\textsf{Symboleo is a formal language used to specify legal contracts.}
\end{mdframed}

\begin{enumerate}
    \item Xtext Grammar Explanation: can be excluded (used in only 2 test cases, as early tests showed that the LLMs do not know Symboleo's language out of the box) or included (in the other 36 test cases).

    \item Modalities and Events: Included, or excluded.

    \item Emotional Prompting: Included, or excluded.

    \item Example Scenarios:
    \begin{itemize}
        \item No example.
        \item Single Scenario: Scenario A.
        \item Pairwise Combinations: Scenarios A+B, A+C, B+A, B+C, C+A, C+B (to check the impact of orderings).
        \item All Scenarios: Scenarios A+B+C.
    \end{itemize}

\end{enumerate}

Finally, each prompt configuration ends with a standardized query for contract generation, with a 4$^{th}$ natural-language contract, this time for a Computer Sale~\cite{parvizimosaed2024symboleopc} that has non-trivial date/event management and business logic:

\begin{mdframed}[style=MyFrame]\small
\textsf{Given the above information, please provide a Symboleo specification (compliant with the language grammar) for the following natural language contract:\\ 
\\
i) The customer orders a computer from a store, to be delivered within 7 days;\\
ii) The customer agrees to pay a deposit worth between 15\% and 20\% of the computer price, on the same day;\\
iii) The customer agrees to pay the remaining amount of the computer price within 10 days of delivery;\\
iv) If delivery is late, the customer has the option (power) to cancel the contract or get a 5\% reduction on the original price and pay within 10 days of delivery.}
\end{mdframed}

\section{Evaluation}
\label{sec:Evaluation}

Each prompt configuration was tested using ChatGPT 4o~\cite{ChatGPT} in ephemeral mode, ensuring that the results from one prompt configuration did not influence subsequent generations. This approach allowed us to isolate the performance of each prompt and configuration independently, preventing data retention from impacting the accuracy of outputs.

In addition, two test cases (\#2 with only scenarios A, B, and C, and \#33 with everything), were repeated across four other LLMs: Claude 3.5 Haiku~\cite{Claude}, Gemini 1.5 Pro 002~\cite{Gemini}, Llama 3.2~\cite{Llama}, and Mistral 7B~\cite{Mistral}. This cross-model testing allows to briefly examine consistency and variability in the generation of \Symboleo/ specifications across different LLMs. Due to Llama's word limit, its prompts had to be divided and processed into three sections, ensuring the entire prompt was presented accurately despite input constraints.

\subsection{Evaluation Metrics}
In our evaluation of the generated \Symboleo/ specifications, we established a set of criteria to assess the quality of each output. Each error type was assigned a weight to reflect its relative impact on the specification's integrity, guiding the prioritization of issues. Errors were classified as high-impact, medium-impact, and low-impact.

\textbf{High-Impact Errors (Weight: 4)}
\begin{itemize}
    
\item\textit{Incorrect Elements Identification: }Misclassifying roles, events, assets, or conditions, such as defining a role as an asset or an event as a role.

\item\textit{Missing Elements Identification: }Omission of essential roles, assets, or events, such as customer or computer. 

\item\textit{Including Information from Outside the Query:} Adding irrelevant elements not specific to the query, such as unrelated roles or assets, detracting from the focus of the specification.

\item\textit{Missing Conditions in the Contract:} Failing to include necessary preconditions, postconditions, or conditions within obligations and powers.

\item\textit{Missing Calculations:} Omitting essential calculations, e.g., total price, from obligations or other contract sections.

\item\textit{Missing All Attributes:} Omitting defining attributes for an entity, such as leaving computer without attributes.

\item\textit{Misunderstanding of Structure Roles:} Placing elements in incorrect sections, e.g., adding a variable in the signature or misplacing an attribute in the obligation section.
\end{itemize}

\textbf{Medium-Impact Errors (Weight: 3)}
\begin{itemize}
\item\textit{Incorrect Data Type Identification:} Assigning incorrect data types to attributes or parameters, such as defining the price as a String instead of a Number.

\item\textit{Inconsistency with the Grammar:} Using invalid constructs that deviate from \Symboleo/'s grammar.

\item\textit{Misidentified Environment Variables: }Using incorrect environment variables altering the contract's intended logic.

\item\textit{Providing Wrong Logic:} Implementing incorrect logical relationships, changing the meaning of obligations or conditions.

\item\textit{Incorrect Calculations:} Errors in calculations, such as miscalculating values or failing to account for parameters.

\item\textit{Including Unnecessary Information:} Adding extra details, such as additional roles or attributes.

\item\textit{Missing Attributes: }Omission of individual attributes (e.g., who := cust).
\end{itemize}

\textbf{Low-Impact Errors (Weight: 2)}
\begin{itemize}
\item\textit{Incorrect Syntax:} Minor syntax errors, such as missing keywords, incorrect delimiters, or misplaced symbols.

\item\textit{Missing Parameters: }Omission of parameters within the signature.
\end{itemize}

\subsection{Results}
By testing our prompt combinations on ChatGPT, we generated 38 distinct \Symboleo/ specifications. We then applied our evaluation metrics, identifying and gathering violations for each structural element of a \Symboleo/ contract, including Domain, Declarations, Preconditions, Postconditions, Obligations, Powers, and Constraints. For each such element, we calculated the total weight/severity of the identified violations. Finally, the overall contract violation weight was determined by summing these total weights across all structural elements.  

\begin{figure}[h!]
        \centering
        \includegraphics[width=1\linewidth]{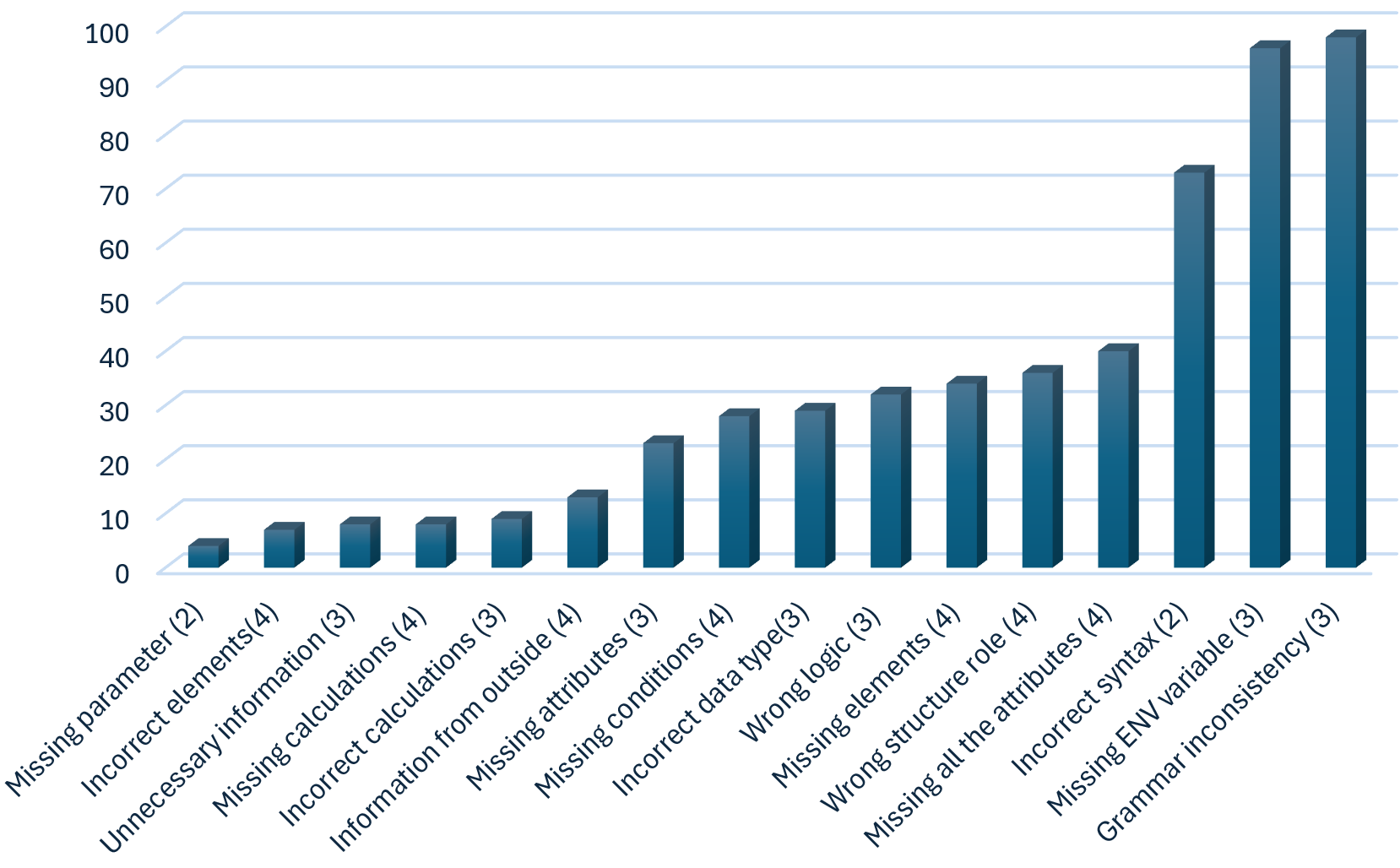}
        \caption{Frequencies of errors across all generated \Symboleo/ specifications}
        \label{fig:enter-label}
    \end{figure}

We calculated the frequency of each error across all generated contracts. Figure~\ref{fig:enter-label} illustrates the distribution of error counts across different error types. Additionally, Table~\ref{ChatGPResults} shows the results of applying our evaluation criteria (considering weights) to each generated \Symboleo/ specification.

\begin{table}[h!]
    \centering
\caption{The generated test cases and the weight of the errors according to each element of \Symboleo/ contract structure. \\
test case= Cas, Contract structure=Con, Domain=Dom, Precondition=Pre, Postcondition=Pos, Signature=Sig, Constraints=Cos, Total error severity=Tot}
\begin{tabular}
{|>{\columncolor{lightgreen}}p{0.85cm}|>{\columncolor{lightgreen}}p{0.31cm}|p{0.31cm}|p{0.31cm}|p{0.31cm}|p{0.31cm}|p{0.31cm}|p{0.31cm}|p{0.31cm}|p{0.31cm}|>{\columncolor{lightgreen}}p{0.31cm}|}
\hline
\rowcolor{lightgreen}
\textbf{Scenario} &\textbf{Cas} & \textbf{Cont} & \textbf{Dom} & \textbf{Dec} & \textbf{Pre} & \textbf{Pos} & \textbf{Sig} & \textbf{OP} & \textbf{Cos} & \textbf{Tot} \\
\hline
No.&1 & 1 & 67 & 67 & 0 & 0 & 9 & 33 & 0 & 177 \\ \hline
ABC&2 & 0 & 14 & 23 & 7 & 0 & 4 & 16 & 0 & 64 \\ \hline
\rowcolor{lightgreen}
        \multicolumn{11}{|l|}{\bf Grammar, no theory, and no prompt}  \\   \hline
No.&3 & 1 & 27 & 83 & 4 & 4 & 9 & 17 & 4 & 149 \\ \hline
A&4 & 0 & 9 & 0 & 0 & 0 & 0 & 15 & 0 & 24 \\ \hline
AB&5 & 0 & 3 & 0 & 0 & 0 & 0 & 3 & 4 & 10 \\ \hline
ABC&6 & 0 & 10 & 3 & 6 & 0 & 4 & 12 & 4 & 39 \\ \hline
AC&7 & 0 & 14 & 6 & 0 & 0 & 7 & 12 & 4 & 43 \\ \hline
BC&8 & 0 & 14 & 3 & 0 & 0 & 4 & 15 & 4 & 40 \\ \hline
BA&9 & 0 & 3 & 0 & 0 & 0 & 0 & 5 & 0 & 8 \\ \hline
CA&10 & 0 & 11 & 3 & 4 & 0 & 4 & 13 & 4 & 39 \\ \hline
CB&11 & 0 & 10 & 19 & 4 & 0 & 0 & 14 & 4 & 51 \\ \hline
\rowcolor{lightgreen} \multicolumn{11}{|l|}{\bf Grammar, no theory, and prompt}  \\   \hline
No.&12 & 0 & 43 & 83 & 0 & 0 & 0 & 32 & 4 & 162 \\ \hline
A&13 & 0 & 3 & 0 & 0 & 0 & 0 & 15 & 0 & 18 \\ \hline
AB&14 & 0 & 3 & 5 & 4 & 0 & 0 & 15 & 0 & 27 \\ \hline
ABC&15 & 0 & 14 & 3 & 4 & 0 & 4 & 10 & 0 & 35 \\ \hline
AC&16 & 0 & 14 & 6 & 4 & 0 & 7 & 7 & 4 & 42 \\ \hline
BC&17 & 0 & 10 & 3 & 4 & 0 & 4 & 7 & 4 & 36 \\ \hline
BA&18 & 0 & 3 & 9 & 0 & 0 & 0 & 8 & 4 & 24 \\ \hline
CA&19 & 0 & 14 & 3 & 4 & 0 & 4 & 10 & 0 & 35 \\ \hline
CB&20 & 0 & 14 & 3 & 4 & 0 & 4 & 10 & 4 & 39 \\ \hline
\rowcolor{lightgreen} \multicolumn{11}{|l|}{\bf Grammar, theory, and no prompt}  \\   \hline
No.&21 & 0 & 43 & 83 & 4 & 4 & 0 & 61 & 4 & 199 \\ \hline
A&22 & 0 & 3 & 0 & 4 & 0 & 0 & 3 & 0 & 10 \\ \hline
AB&23 & 0 & 3 & 0 & 4 & 0 & 0 & 3 & 0 & 10 \\ \hline
ABC&24 & 0 & 14 & 3 & 4 & 0 & 0 & 16 & 4 & 41 \\ \hline
AC&25 & 0 & 14 & 3 & 4 & 0 & 4 & 14 & 4 & 43 \\ \hline
BC&26 & 0 & 14 & 3 & 0 & 0 & 4 & 14 & 4 & 39 \\ \hline
BA&27 & 0 & 3 & 0 & 0 & 0 & 0 & 8 & 4 & 15 \\ \hline
CA&28 & 0 & 14 & 3 & 4 & 0 & 4 & 12 & 4 & 41 \\ \hline
CB&29 & 0 & 14 & 7 & 0 & 0 & 4 & 10 & 4 & 39 \\ \hline
\rowcolor{lightgreen} \multicolumn{11}{|l|}{\bf Grammar, theory, and prompt}  \\   \hline
No.&30 & 4 & 35 & 83 & 7 & 7 & 0 & 28 & 7 & 171 \\ \hline
A&31 & 0 & 3 & 6 & 0 & 0 & 0 & 7 & 0 & 16 \\ \hline
AB&32 & 0 & 7 & 0 & 0 & 0 & 0 & 3 & 0 & 10 \\ \hline
ABC&33 & 0 & 14 & 3 & 4 & 0 & 4 & 10 & 4 & 39 \\ \hline
AC&34 & 0 & 6 & 9 & 0 & 0 & 0 & 10 & 4 & 29 \\ \hline
BC&35 & 0 & 10 & 3 & 0 & 0 & 3 & 10 & 4 & 30 \\ \hline
BA&36 & 0 & 7 & 2 & 0 & 0 & 0 & 8 & 0 & 17 \\ \hline
CA&37 & 0 & 14 & 6 & 4 & 0 & 7 & 10 & 4 & 45 \\ \hline
CB&38 & 0 & 12 & 3 & 4 & 0 & 4 & 14 & 4 &  41 \\
\hline
\rowcolor{lightgreen}
 \multicolumn{2}{|c|}{\textbf{Total}} & \textbf{6} & \textbf{534} & \textbf{539} & \textbf{92} & \textbf{15} & \textbf{98} & \textbf{510} & \textbf{103} &  \\
\hline
\end{tabular}
\label{ChatGPResults}
\end{table}   

\textbf{Impact of Context Changes}: We examined how different modifications to the input data or prompt configurations influence the model's results. 

\textit{Impact of Grammar and Examples}: To evaluate the impact of grammar and examples, we analyzed the test cases based on input categories that either included or omitted grammar rules and/or theoretical guidance. Without examples (cases 1, 3, 12, 21, and 30), ChatGPT tends to produce many more errors compared to when examples are included. This trend is clearly demonstrated in case 1, where the grammar rules and examples were not provided and the total error weight was 177. However, by providing three examples (case 2), the error weight dropped substantially to 64.
Interestingly, the severity of errors increased when only grammar rules were provided without examples, as seen in case 21. The error count was higher than in case 1, where no guidance was given beyond the initial query. This is because providing the grammar alone prompted ChatGPT to generate more complex code, leading to more errors due to incorrect or incomplete application of the rules. This suggests that while grammar guidance alone can increase the volume of generated code, it may also elevate the risk of syntax misunderstandings, grammar misinterpretations, or logic errors if not accompanied by concrete examples.

\textit{Context Sensitivity}: When using scenarios A and B together or just scenario A alone, the resulting \Symboleo/ specifications had fewer errors. However, including scenario C in the configuration noticeably increased the number of errors. This is because scenario C does not include any environment variables (ENV) in the domain, declaration, obligation, and power sections, unlike scenarios A and B. Consequently, when scenario C was introduced, ChatGPT tended to \textit{omit} environment variables altogether, even though they had been identified and utilized in the more detailed examples provided by A and B. The lack of environment variable references in scenario C caused ChatGPT to deprioritize or ignore them, leading to an increase in related errors in the generated specifications.
The model adapts to recent patterns when the context changes, even at the expense of previously established elements, highlighting its sensitivity to input context.

\textit{Impact of the Prompt}: Given ChatGPT's high sensitivity to the input context, it is crucial to isolate the effect of examples on the generated specifications to accurately evaluate the impact of the prompt itself. This helps avoid generating varied outputs that could unintentionally introduce or correct errors. We focus on the grammar and theory with/without emotional prompt categories  since the inclusion of theory helps reduce ChatGPT's dependence on complex or incomplete grammar rule applications, instead offering more consistent and straightforward guidance. In this analysis, cases 21 and 30 serve as ideal comparisons due to their alignment within these requirements. By disregarding errors that typically arise or are corrected when testing the same input multiple times (see upcoming ``Variations in Output with Same Input'' section), we can shift our focus to the size of the generated code, the quality of the solution, and the main issues that are fixed only when providing examples. Notably, the code size in case 30 (75 lines) is 87\% larger than that in case 21 (40 lines). Although one of the primary issues in case 21 was a misinterpretation of the environment variable, this problem persisted in case 30. However, case 30 demonstrated a more complete and refined logic by incorporating additional variables and conditions, and resolving incomplete obligations that were present in case 21.

This observation led us to compare cases 3 and 12 to investigate differences in code size, quality, and the types of errors generated across these input categories (grammar only, without theory, with/without emotional prompts). Notably, case 3 produced a larger code size than case 12 by around seven code lines, but this was mainly due to unnecessary additions like unnecessary end markers for sections (e.g., \textsf{endprecondition}, \textsf{endpostcondition}). In contrast, case 12 showed major improvements in code quality. For example, instead of generic labels like \textsf{O1}, \textsf{P1}, etc., the obligations and powers were named more descriptively, such as \textsf{O\_PayDeposit} and \textsf{P\_LateDeliveryReduction}. This enhanced the clarity and readability of the generated specifications. Moreover, case 12 combined some obligations and included additional conditions and calculations, showcasing a more comprehensive understanding of the contract logic. However, similar to the issues observed in cases 21 and 30, the model still struggled with identifying and correctly using environmental variables, despite noticeable overall enhancements in code structure and logic.    

\textbf{Variations in Output with Same Input}: When we provided the same test file to ChatGPT multiple times, it generated different \Symboleo/ specifications. The variations included changes such as using constants as parameters or variables and moving conditions from obligations to the declaration section. While these adjustments sometimes corrected or introduced minor errors, the resulting error weight differences could reach up to 8 points. Therefore, we set a margin of error at 8; any two specifications with the same design category and a weight difference of 8 or less are considered equivalent. Applying this error margin to the cases in Table~\ref{ChatGPResults}, we observed that tests using the same examples, even when presented in a different order, generally produced similar \Symboleo/ specifications. For instance, cases like 5 and 9, 8 and 10, as well as 23 and 27, showed comparable results, indicating that the variations in example order had minimal impact on the overall structure and quality of the generated specifications.

\textbf{Variation of Cases Across Input Categories}: In contrast, when comparing test cases across different input categories, we found that the newly introduced errors were entirely distinct from those generated when providing the same cases without variations. Including theoretical guidance with the input greatly improved the quality of the generated code, particularly by reducing grammar-related errors. However, this approach also introduced a new set of errors associated with the application of the added theoretical concepts. 
For instance, when analyzing cases 15 and 24, we observed that although the total number of errors was nearly the same, the types of errors differed. In case 24, complex expressions stemming from incomplete grammar rule applications were largely eliminated due to the explicit inclusion of semantic explanations. Yet, as expected, new errors emerged unrelated to the grammar, such as issues with the implementation of the theoretical concepts and incorrect application of domain-specific logic. This shift highlights the trade-off between reducing syntax errors and introducing new, semantic-specific errors when additional information is incorporated into the input.

\textbf{Error Distribution}: In this section, we will discuss how errors are distributed across various error types within all input categories. As shown in Fig.~\ref{fig:enter-label}, the most significant challenges ChatGPT encounters are adhering to the grammar, correctly identifying environment variables, and maintaining correct syntax. They composed 49\% of the total number of violated metrics and are in a range between 70 to 100 errors for each. An example of inconsistency with the grammar observed in the generated cases is as follows:
\begin{lstlisting}[language=symboleo, basicstyle=\footnotesize\ttfamily, numbers=none,  columns=fullflexible]
Powers
 reducePrice: Happens(Violated(obligations.delivery)) -> P(customer, store, true, Happens(priceReduced) and Happens(paidAfterReduction));
\end{lstlisting}

According to the grammar rules, the consequent of a power must be a power function (e.g., to suspend or trigger obligations, or to terminate contracts). Thus, the correction involves removing the condition from the power and instead triggering the \textsf{oReducePrice} obligation, which handles price reductions. 
The correct code is:
\newpage
\begin{lstlisting}[language=symboleo, basicstyle=\footnotesize\ttfamily, numbers=none,  columns=fullflexible]
Powers
  reducePrice: Happens(Violated(obligations.delivery))-> P(customer, store, true, Triggered(obligations.oReducePrice) );
\end{lstlisting}

Errors with medium occurrence, ranging from 20 to 40 instances, make up 41\% of the total and include missing attributes, elements, and conditions, as well as incorrect logic, misidentified structural roles, and incorrect data type detection. Here is an example of incorrect data type identification: 
\begin{lstlisting}[language=symboleo, basicstyle=\footnotesize\ttfamily, numbers=none,  columns=fullflexible]
Declarations
        deposit: Number;
        remainingPayment: Number;
        deliveryDate: Date;
        lateDelivery: Boolean := false;
\end{lstlisting}

The correction involves updating the data types to events and eliminating \textsf{lateDelivery} and \textsf{deliveryDate}, which were initially conditions within obligations and powers. 
It is important to note that \textsf{Paid} is a domain data type and should be defined as an event in the domain section, together with an amount attribute that will be dynamically provided at runtime as an environment variable. The corrected code is as follows: 
\begin{lstlisting}[language=symboleo, basicstyle=\footnotesize\ttfamily, numbers=none,  columns=fullflexible]
Domain computerPurchaseDomain
Paid: isAn Event with Env amount: Number; 
endDomain
Declarations
        deposit:Paid;
        remainingPayment:Paid;
\end{lstlisting}

In contrast, the least frequent errors, representing only 10\% of the total, occur 4 to 13 times and include issues such as irrelevant information, unnecessary details, incorrect calculations, missing calculations, incorrect elements identification, missing parameters, and incorrect document structure. New information extracted from the examples is often added to the domain and declarations sections. In the following example, additional attributes such as \textsf{name} and \textsf{location} were included, along with a \textsf{Currency} enumeration, even though these details were not specified in the original contract description. 
\begin{lstlisting}[language=symboleo, basicstyle=\footnotesize\ttfamily, numbers=none,  columns=fullflexible]
Domain computerSaleDomain
  Customer isA Role with name: String;
  Store isA Role with location: String;
  Currency isAn Enumeration(USD, EUR, CAD);
\end{lstlisting}

The correction involves removing the unnecessary elements.
\subsection{Other LLMs}
To compare the performance of other LLMs, two test cases were reused: one with minimal guidance (case 2) and another with comprehensive input, including examples, grammar rules, theory, and an emotional prompt (case 33). Table~\ref{LLMsResults} summarizes the results, categorizing and weighing errors across contract sections. The Llama model showed high error rates with minimal input but excelled in case 2 with detailed guidance. Mistral maintained consistent low error severity across both scenarios, identifying but failing to declare environment variables in the domain section. Gemini showed major improvement with more input. From a readability perspective, Llama and Claude produced poorly organized documents, lacking proper indentation and comments. ChatGPT's performance improved notably with more input, reducing weighted errors from 64 to 39. Note that, on our metrics, all four LLMs have scored better than ChatGPT, especially Mistral.
\begin{table}[t]
\centering
\caption{Metric evaluation results for four other LLMs.}
\begin{tabular}
{|>{\columncolor{lightgreen}}p{1cm}|p{0.38cm}|p{0.38cm}|p{0.35cm}|p{0.35cm}|p{0.35cm}|p{0.35cm}|p{0.35cm}|p{0.35cm}|>{\columncolor{lightgreen}}p{0.35cm}|}
\hline
\rowcolor{lightgreen}
\textbf{LLM}  & \textbf{Cont} & \textbf{Dom} & \textbf{Dec} & \textbf{Pre} & \textbf{Pos} & \textbf{Sig} & \textbf{OP} & \textbf{Cos} & \textbf{Tot} \\
\hline
\rowcolor{lightgreen}
        \multicolumn{10}{|l|}{\bf No Grammar, No theory, and No prompt -- Test Case 2}  \\   \hline
Claude& 0&	6&	9&	0&	0&	0&	18&	7&	40
\\ \hline
Gemini &0	&0&	18&	0&	0&	3&	17&	8&	46
 \\ \hline
Llama & 0	&3&	14&	7&	0&	3&	20&	7&	54  \\ \hline
Mistral & 0&	0&	3&	0&	0&	3&	18&	7&	31  \\ \hline
\rowcolor{lightgreen}
        \multicolumn{10}{|l|}{\bf With grammar, theory, and prompt -- Test Case 33}  \\   \hline
Claude & 0&	3&	3&	0&	0&	0&	22&	7&	35 \\ \hline
Gemini & 0&	11&	14&	0&	0&	0&	9&	0&	34 \\ \hline
Llama & 0&	7&	7&	0&	0&	0&	13&	0&	27 \\ \hline
Mistral & 0&	3&	6&	0&	0&	3&	11&	7&	30
 \\ \hline
\end{tabular}
\label{LLMsResults}
\end{table}

\section{Discussion}
\label{sec:Discussion}
This study explores the use of LLMs to formalize natural-language legal contracts by generating \Symboleo/ specifications. It demonstrates the potential of LLMs to convert legal documents into machine-readable formats while identifying areas for further improvement.
\subsection{Implications}
\begin{itemize}
    \item Impact of the context changes: The model's sensitivity to context changes can be mitigated in two ways. First, by revising the provided information dynamically, and second, by running each test multiple times to gain a statistically-significant understanding of the primary considerations, distinguishing major findings from minor variations that can be overlooked or adjusted.

    \item Consistency and quality trade-off: For more consistent results with reduced variability across generated cases, adjusting the LLM's temperature parameter could help. Lowering the temperature decreases a model's creativity, possibly making it adhere more strictly to grammar rules. However, this approach may limit the benefits gained from concrete examples and detailed input. This could also lead to overly complex solutions as the model rigidly applies grammar rules without practical guidance, relying solely on its interpretation of the grammar instructions.

    \item Need for guidance: without precise guidance such as clear grammar rules, detailed descriptions, and accurate concrete examples, the model is prone to generating incorrect specifications. Critical errors like missing elements or conditions can significantly impact the accuracy of legal contract monitoring. 

    \item Need for experience: While the model offers considerable time savings by automatically drafting specifications from natural language inputs, the involvement of domain experts remains essential. Given the complexity and volume of the generated specifications, beginners may struggle to identify logical flaws or detect missing elements, underscoring the need for expert oversight in the process.

    \item Impact of emotional prompting: While such prompts can add depth and engagement to the result explanations, the technical quality impact observed was not really significant and was even at times counterproductive.
\end{itemize}

\subsection{Limitations}
This study has several limitations that impact its findings. First, the generated \Symboleo/ specifications were not evaluated using automated tools or formal validation methods, instead relying on manual (and possibly biased) evaluation of the LLM output. This lack of automated evaluation may have allowed errors in syntax, logic, or compliance to go unnoticed, affecting the evaluation scores. To mitigate this threat to validity, the first three authors collaborated on the assessment of the specifications. 

Second, the evaluation metrics used are very specific to \Symboleo/, were not validated by others, and are not easily generalizable to other DSLs.

Third, given that this is a first \Symboleo/-oriented experiment, very few types of prompt components were used, fine-tuning was ignored, and the tests themselves were not systematically repeated multiple times to achieve statistical significance. Additionally, we used a limited number of contracts (all in English only) as examples, and all our test cases aimed to generate a specification for the same small contract description. Addressing these aspects is left to future work.

Lastly, reliance on a few Web-based versions of LLMs may have negatively affected the quality of the outputs. More advanced or contract-friendly models could be explored, also via their APIs, which usually provide more flexibility than Web-based chat interfaces.

\section{Related Work}
\label{sec:RelatedWork}
There is much recent activity related to the exploration of LLMs for various model/DSL generation tasks. Lamas et al.~\cite{Lamas2024DSLgen} have recently looked into the use of LLMs to support code generation targeting little-known DSLs. They proposed the \textit{DSL-Xpert} tool to provide semantic parsing as a means to improve code generation reliability. As in our experiment, they also use the DSL grammar and few-shot learning in their approach. That tool could be further investigated in \Symboleo/'s context. Bassamzadeh and Methani~\cite{bassamzadeh2024DSLcodeGen} compared LLM fine-tuning and Retrieval Augmented Generation (RAG~\cite{Fan2024RAGsurvey}) for DSL-oriented code generation, with an emphasis on unseen API function names. The simpler RAG approach is shown to match fine-tuning in avoiding LLM hallucinations in that context.

Others have also explored the generation of software engineering models from natural-language requirements (e.g., to UML sequence diagrams~\cite{Ferrari2024-SeqDiagrams}), domain descriptions (e.g., to UML class diagrams~\cite{Yang2024-MultiStep}), and project descriptions (e.g., to goal models~\cite{chen2023gpt4goalmodels}). Mosthaf et al.~\cite{Mosthaf2024-DSLassistant} have investigated the use of LLMs as means to generate and evolve grammars and DSLs themselves, which is different from our objectives (as \Symboleo/ already exists). Ma et al.~\cite{ma2024SpecGen} have used LLMs for inferring formal specifications from existing Java programs, which is the opposite problem. 

For smart contracts, Napoli et al.~\cite{Napoli2024AutomaticSCgen} used an LLM-based pipeline to assist the generation of secure smart contract code, but not directly from legal contracts. Leite et al.~\cite{Leite2024} used ChatGPT for developing design-by-contract specifications from natural language that can be used to model-check smart contracts written in Solidity. Their source documents are however Ethereum standards, not legal contracts.

For \Symboleo/, Meloche et al.'s work~\cite{meloche2023towards,Meloche24,SymboleoWebaper2024} on a controlled natural language (CNL) empirically constructed from existing legal contracts is reliable, but limited in scope. A  \Symboleo/ specification must first exist for a contract template, and then NLP/CNL-based refinements to the template at predetermined locations are automatically transformed to corresponding modifications to the original specification. Our research offers higher automation capabilities through the use of LLMs, which can interpret and translate a broader range of contractual concepts (albeit likely with more errors) into \Symboleo/ specifications, without initial manual input.

\section{Conclusion and Future Work}
\label{sec:Conclusion}

This paper explored how to automate the generation of formal specification from natural-language legal contracts using LLMs. Targeting the \Symboleo/ DSL, known for its formal rigor, the research highlights LLMs' potential to reduce the manual effort of translating English contracts into machine-readable formats. Testing 38 prompt configurations across multiple LLMs revealed promising results, even for a barely known DSL such as \Symboleo/ and a few relatively naïve prompting techniques. Our documented results and implications partially answer our research question (\textbf{RQ}). However, challenges remain, including syntax inaccuracies, semantic inconsistencies, and difficulties in capturing the complex logic of legal contracts, indicating the need for further refinement and improvement.

Future work aims to address these challenges through automated validation and error-detection mechanisms to improve correctness and compliance. Exploiting LLM-level parameters, using additional and established prompt-engineering techniques (e.g., Chain-of-Thought), and exploring other translation examples are all options that deserve further investigation. Fine-tuning LLMs with contractual and \Symboleo/-specific datasets could also improve domain-specific understanding.

This preliminary experiment tested GPT-4o with 38 prompt configurations, but it used only two test cases with the other LLMs, which were mostly free Web-based versions. A comprehensive evaluation of many test cases across diverse LLMs is necessary to assess effectiveness. Involving real legal experts in a more complex experiment would also help understand the efficiency and adoptability of an LLM-based approach.

Advancing the integration of LLMs and formal specification generation has the potential to revolutionize legal contract specification, validation, and execution, eventually paving the way for widespread adoption of smart legal contracts.

\bibliographystyle{IEEEtran}
\bibliography{RAISE2025-Symboleo-LLM}

\end{document}